\documentclass[pra,twocolumn,floatfix,a4paper,superscriptaddress]{revtex4}

\usepackage{amssymb}                        
\usepackage{bm} 
\usepackage{color,amsmath,txfonts}
\usepackage{graphicx}
\usepackage{siunitx}
\usepackage{subfigure}
\usepackage{verbatim}
\usepackage{dcolumn}
\usepackage{bm}
\usepackage{epsf}
\usepackage{xcolor}
\usepackage{hyperref}
\usepackage{hhline}
\usepackage{float}
\usepackage{enumerate}
\usepackage{bbm}
\usepackage{lipsum}
\usepackage{mathrsfs}
\usepackage{ulem}
\usepackage{natbib}
\usepackage{leftindex}

\begin{document}
	
	\title{Generation of mechanical cat-like states via optomagnomechanics}
	
	\author{Hao-Tian Li}
	\altaffiliation{These authors contributed equally to this work}
	\affiliation{Zhejiang Key Laboratory of Micro-Nano Quantum Chips and Quantum Control, and School of Physics, Zhejiang University, Hangzhou 310027, China}
	\author{Hong-Bin Wang}
	\altaffiliation{These authors contributed equally to this work}
	\affiliation{School of Physics and Shing-Tung Yau Center, Southeast University, Nanjing 210096, China}
	\author{Zi-Xu Lu}
	\affiliation{Zhejiang Key Laboratory of Micro-Nano Quantum Chips and Quantum Control, and School of Physics, Zhejiang University, Hangzhou 310027, China}
	\author{Jie Li}\thanks{jieli007@zju.edu.cn}
	\affiliation{Zhejiang Key Laboratory of Micro-Nano Quantum Chips and Quantum Control, and School of Physics, Zhejiang University, Hangzhou 310027, China}

	\begin{abstract}
		We propose an optomagnomechanical approach for preparing a cat-like superposition state of mechanical motion. Our protocol consists of two steps and is based on the magnomechanical system where the magnetostrictively induced displacement further couples to an optical cavity mode via radiation pressure. We first prepare a squeezed mechanical state by driving the magnomechanical system with two microwave pulses. We then switch off the microwave drives and send a weak red-detuned optical pulse to the optical cavity to weakly activate the optomechanical anti-Stokes scattering.  We show that $k$ phonons can be subtracted from the prepared squeezed thermal state, conditioned on the detection of $k$ anti-Stokes photons from the cavity output field, which prepares the mechanical motion in a cat-like state. The work provides a new avenue for preparing mechanical superposition states by combining opto- and magnomechanics and may find applications in the study of macroscopic quantum states and the test of collapse theories.	
	\end{abstract}
	
	\maketitle

	\section{Introduction}

	One distinctive feature of quantum mechanics is embodied by the fact that it allows the linear superposition of quantum states.  As one special type of superposition states, the cat state refers to a quantum state composed of a superposition of two coherent states with opposite phases. 
	Cat states were routinely generated in a variety of microscopic systems, including trapped ions \cite{CM96,DL05}, microwave photons \cite{BV13} optical photons \cite{KH15,AO06,JSN06}, atomic ensembles \cite{AO19}, and hybrid atom-light systems \cite{BH19}. However, preparing such states in macroscopic systems, e.g., a massive mechanical oscillator, is extremely challenging. Impressively, a recent experiment has demonstrated the so-far largest cat state of a bulk acoustic-wave resonator via the manipulation of a superconducting qubit \cite{MB23}.  Apart from this system, proposals \cite{IS20,Tan20} indicate that the optomechancial system \cite{MA14} is also a promising system to prepare cat-like states of a macroscopic mechanical oscillator. 
	
	In recent years, the magnomechanical system \cite{Zuo} has attracted much attention due to its potential for preparing macroscopic quantum states \cite{JL18,JL19b}, quantum sensing~\cite{Jing24}, and its rich nonlinearities, such as magnon-phonon cross-Kerr effect \cite{RCS22} and magnonic frequency combs \cite{Xiong23,Dong23}. For a large-size magnetic sample, e.g., an yttrium-iron-garnet (YIG) sphere, the magnetostrictive interaction is a dispersive type, which couples magnon excitations to the deformation displacement. Further coupling the magnomechanical displacement to an optical cavity via radiation pressure forms a hybrid optomagnomechanical (OMM) system \cite{Fan,Dong22,QST23}, which has been proven to be a promising system to realize the optical readout of magnonic states \cite{QST23} and generate microwave-optics \cite{LPR23,LPR25,Xiong25,Zhang25} and magnon-atom entanglement \cite{FanPRA23,Di24}.
	
	Here, we present a scheme for preparing a mechanical cat-like state based on the OMM system {using fast microwave and optical pulses}. We adopt the approach of subtracting excitations from the squeezed vacuum state \cite{MD97,AB07,TJ16}.  Specifically, our scheme consists of two steps: $i$) Preparing a mechanical squeezed state in the magnomechanical subsystem by applying {a pair of red- and blue-detuned microwave pulses} to the magnon mode; $ii$) Subtracting phonons from the generated squeezed state by sending a weak red-detuned optical pulse to the optomechanical cavity. The detection of $k$ anti-Stokes photons in the cavity output field heralds a  $k$-phonon-subtracted squeezed state, corresponding to a cat-like state of mechanical motion.
	
	The paper is organized as follows. In Sec.~\uppercase\expandafter{\romannumeral2}, we introduce the OMM system and {the pulse sequence used in our protocol. 
		We then show how to generate a mechanical squeezed state in the magnomechanical subsystem by applying two microwave pulses in Sec.~\uppercase\expandafter{\romannumeral3}, and how to subtract $k$ phonons from the mechanical squeezed state in the optomechanical subsystem by applying a red-detuned optical pulse in Sec.~\uppercase\expandafter{\romannumeral4}.  }   Finally, we summarize our findings in Sec.~\uppercase\expandafter{\romannumeral5}.

	\section{The OMM system and the pulse sequence}
	\label{model}

	\begin{figure}
		\centering
		\includegraphics[width=1\linewidth]{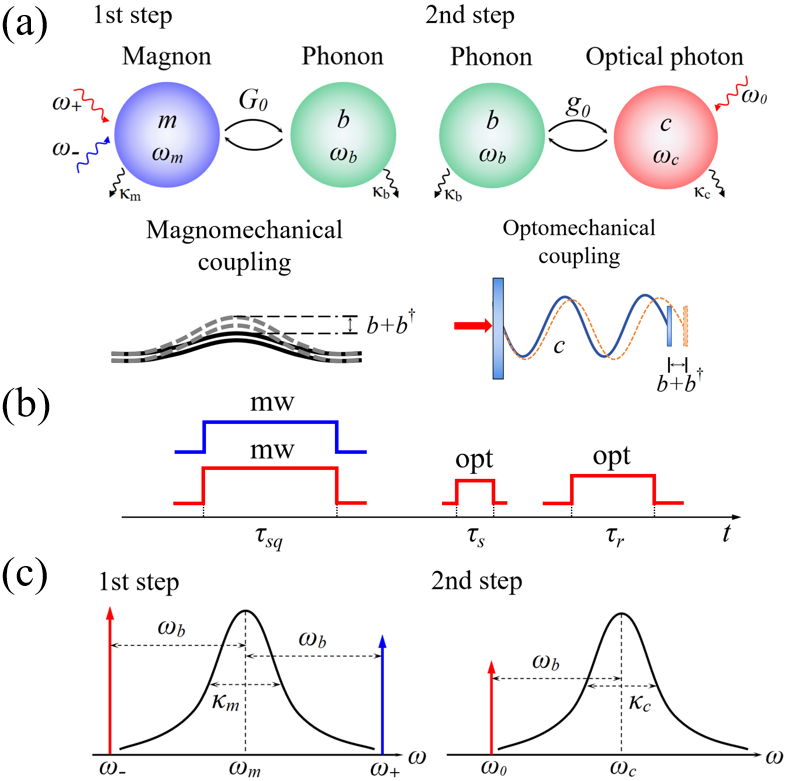}
		\caption{(a) Two-step protocol adopted to generate cat-like states of mechanical motion in an OMM system. First step: the magnon mode is driven by two microwave pulses at frequencies $\omega_\pm=\omega_m \pm \omega_b$ to prepare a mechanical squeezed state. Second step: a weak red-detuned optical pulse is sent to the optomechanical cavity to subtract phonons, conditioned on the detection of anti-Stokes photons in the cavity output field. {(b) The pulse sequence used in our protocol. Two microwave pulses at frequencies $\omega_m \pm \omega_b$ and with duration $\tau_{\rm sq}$ are used to prepare a mechanical squeezed state.  A red-detuned optical pulse with duration $\tau_s$ is sent to the cavity to subtract phonons from the squeezed state, conditioned on the detection of anti-Stokes photons. Once a cat-like state is generated, another red-detuned optical pulse with duration $\tau_r$ is sent to read out the mechanical state for subsequent tomography operation.} (c) Frequencies of the magnomechanical subsystem and two microwave driving fields in the first step, and of the optomechanical subsystem and the weak optical pulse in the second step.}
		\label{fig1}
	\end{figure}

	The OMM system consists of a magnon mode (e.g., the Kittel mode \cite{Kittel}) and an optical cavity mode, and both couple to a mechanical mode via the magnetostrictive and radiation-pressure interactions, respectively, as depicted in Fig.~\ref{fig1}(a). The magnomechanical system can be a YIG micro bridge structure~\cite{PRAp}, and the optical cavity can be formed by attaching a small highly-reflective mirror pad to the surface of the micro bridge~\cite{LPR23,SG}. Another promising system could be the opto- and magnomechanical systems coupled via direct physical contact \cite{Dong22}.
	The YIG crystal is placed inside a uniform bias magnetic field and further driven by a microwave field applied via, e.g., a loop antenna. The magnon mode is then activated and coupled to the mechanical displacement via the magnetostrictive interaction, and the latter further couples to the optical cavity via radiation pressure. We consider the situation where the mechanical resonance frequency is much lower than the magnon frequency, such that the magnomechanical coupling is a dispersive type \cite{Zuo}.  The Hamiltonian of the three-mode OMM system reads
	\begin{equation}
		H/\hbar=\sum_{j=m,b,c} \omega_j j^\dagger j + \left(G_0m^\dagger m -g_0c^\dagger c \right)  \left(b+b^\dagger \right)  +H_{\rm dri},
	\end{equation}
	where $j$ ($j^\dagger$), $j=m,b,c$, are the annihilation (creation) operators of the magnon, mechanical, optical cavity modes, respectively, and $\omega_j$ are their corresponding frequencies. $G_0$ ($g_0$) is the bare magnomechanical (optomechanical) coupling strength, and $H_{\rm dri}$ represents the driving Hamiltonian, which is different and will be specified  in the following two steps.
	
	{Our protocol adopts a series of fast microwave and optical pulses, as shown in Fig.~\ref{fig1}(b).  In the first step, two microwave pulses with different frequencies and duration $\tau_{\rm sq}$ are used to squeeze the mechanical motion~\cite{Schwab15}. Once a squeezed state is prepared and after a short period when magnons die out, a red-detuned optical pulse with duration $\tau_{s}$ is sent to the optical cavity to subtract phonons from the squeezed state.  After anti-Stokes photons being detected from the cavity output field, which heralds a mechanical cat-like state, another red-detuned optical pulse with duration $\tau_{r}$ is sent to map the mechanical state to the cavity output field~\cite{Jie18a,Tan20} for subsequent tomography of the state. }

	\section{Magnomechanical SQUEEZING OF MECHANICAL motion}
	\label{sqzing}

	{In the first step, to simplify the model we consider two flattop microwave pulses~\cite{Jie18a}. During the pulse, the driving strength can be assumed constant, which corresponds to a constant Rabi frequency introduced below. After a short period $\tau_{\rm sq}$, when the mechanical system evolves into a stationary state with an invariant degree of squeezing, we then switch off the microwave pulses.} Due to the absence of a strong laser driving field to enhance the optomechanical coupling, the effective optomechanical coupling is very weak. The optical cavity is essentially decoupled from the driven magnomechanical system. Therefore, the tripartite OMM system reduces to an effective two-mode system with the Hamiltonian given by
	\begin{align}
		\begin{split}
			H_{\rm MM}/\hbar&=\omega_mm^\dagger m+\omega_bb^\dagger b+G_0m^\dagger m \left(b+b^\dagger \right)\\
			&+\left[ \left(\Omega_+e^{-i\omega_+t}+\Omega_-e^{-i\omega_-t} \right)m^\dagger - \mathrm{H}. \mathrm{c}. \right],
		\end{split}
	\end{align}
	which indicates that the magnon mode is driven by two microwave fields at the frequencies $\omega_\pm=\omega_m\pm\omega_b$, and $\Omega_\pm$ denote the associated magnon-drive coupling strength, i.e., the Rabi frequency. {For a YIG micro bridge, the Rabi frequency is defined as $\Omega=\frac{\sqrt{5}}{4}\gamma\sqrt{N}B_0$~\cite{JL18}, where the gyromagnetic ratio $\gamma/2\pi=28$ GHz/T, the total number of spins $N=\rho V$ with $V$ being the volume of the micro bridge and $\rho=4.22\times10^{27}$m$^{-3}$ being the spin density of the YIG, and $B_0$ is the amplitude of the driving magnetic field. The drive amplitude is related to the drive power via $B_0=\sqrt{2\mu_0P_0/(lwc)}$~\cite{QST23}, where $\mu_0$ is the vacuum magnetic permeability, $c$ is the speed of the electromagnetic wave propagating in vacuum, and $l$ and $w$ are the length and width of the micro bridge, which is approximated as a cube for simplicity. } 

Incorporating the dissipation and input noise terms, we obtain the following quantum Langevin equations (QLEs) of the magnomechanical system:
\begin{align}
	\begin{split}
		\dot{m}&=\left(-i\omega_{m}-\frac{\kappa_{m}}{2} \right)m-iG_{0}m \left(b+b^{\dagger} \right)\\&-i \left(\Omega_+e^{-i\omega_+t}+\Omega_-e^{-i\omega_-t} \right)+\sqrt{\kappa_{m}}m_{in},
		\\\dot{b}&=\left(-i\omega_{b}-\frac{\kappa_{b}}{2} \right)b-iG_{0}m^{\dagger}m+\sqrt{\kappa_{b}}b_{in},
	\end{split}
\end{align}
where $\kappa_m$ and $\kappa_b$ are the dissipation rates of the magnon and mechanical modes, respectively, and $O_{in}$ ($O = m, b$) denote the input noises of the two modes, and their nonzero correlation functions are $\langle O_{in}^{\dagger}(t)O_{in}(t')\rangle=N_O(\omega_O)\delta(t-t')$ and $\langle O_{in}(t)O_{in}^{\dagger}(t')\rangle=\left[N_O(\omega_O)+1 \right]\delta(t-t')$, with $N_O(\omega_O)= \left[ {\rm exp}(\hbar\omega_{O}/k_{B}T) - 1 \right]^{-1}$ being the equilibrium mean thermal magnon/phonon number and $T$ being the bath temperature.

Following the standard linearization treatment, we write each mode operator $O$ as a large classical average $O_s$ plus a small fluctuation operator $\delta O$, i.e., $O = O_s + \delta O$ ($O = m, b$). Substituting them into the above QLEs, the equations are separated into two sets: one for the classical averages and the other for the quantum fluctuations. The two strong driving fields at frequencies $\omega_\pm$ lead the magnon mode to be dominant at the two drive frequencies, which allows us to assume that $m_s \simeq m_+ e^{-i\omega_+t} + m_- e^{-i\omega_-t}$ \cite{YDW13,HT13,JL15}. As $G_0$ is very weak~\cite{Zuo}, and typically $G_0 |$Re$ \,b_s|\ll \omega_b$, we can safely neglect the small term $iG_0m_s(b_s+b_s^{\ast})$ in getting the solutions of the averages $m_\pm$, which are obtained as	
\begin{align}
	\begin{split}
		m_\pm = \frac{\Omega_\pm}{\omega_\pm - \omega_m + i \frac{\kappa_m}{2}}.
	\end{split}
\end{align}

We also obtain the linearized QLEs for the quantum fluctuations by neglecting small second-order fluctuation terms, which, in the interaction picture with respect to $\hbar \omega_m m^\dagger m + \hbar \omega_b b^\dagger b$, are given by

\begin{align}\label{QQQ}
	\begin{split}
		\delta \dot{b} &= -\frac{\kappa_b}{2}\delta b - i \left(G_- + G_+ e^{2i\omega_bt} \right)\delta m \\
		&- i \left(G_+ + G_-e^{2i\omega_bt} \right)\delta m^\dagger + \sqrt{\kappa_b}b_{in},\\
		\delta \dot{m} &=-\frac{\kappa_m}{2}\delta m - i \left(G_- + G_+e^{-2i\omega_bt} \right)\delta b \\
		&- i \left(G_+ + G_-e^{2i\omega_bt} \right)\delta b^\dagger + \sqrt{\kappa_m}m_{in}.
	\end{split}
\end{align}
Here, $G_\pm = G_0m_\pm$ are the enhanced magnomechanical coupling strengths due to the two strong driving fields at frequencies $\omega_\pm$, which can be set real by adjusting the phases of the drive fields to have real $m_\pm$. 
Due to the time-dependent fast oscillating terms in Eq.~\eqref{QQQ}, the above QLEs are difficult to solve. However, under the condition of $\kappa_b, \kappa_m, G_\pm \ll \omega_b$ \cite{JL15}, the rotating-wave approximation (RWA) can be made by neglecting the fast oscillating terms. Consequently, we obtain
\begin{align}\label{qqqq}
	\begin{split}
		\delta \dot{b}&=-\frac{\kappa_b}{2}\delta b-i \left(G_-\delta m +G_+\delta m^\dagger \right)+\sqrt{\kappa_b}b_{in},\\ 
		\delta \dot{m}&=-\frac{\kappa_m}{2}\delta m-i \left(G_-\delta b+G_+\delta b^\dagger \right)+\sqrt{\kappa_m}m_{in},
	\end{split}
\end{align}
which indicates that the drive field at $\omega_-$ is responsible for cooling the mechanical motion by activating the magnomechanical beamsplitter-type interaction (anti-Stokes process), while the drive field at $\omega_+$ activates the parametric down-conversion interaction (Stokes process). The proper combination of these two interactions can induce a squeezed mechanical mode, which we show later.  To maintain the system stability, the strength of the anti-Stokes scattering should be stronger than that of the Stokes scattering, i.e., $G_- > G_+$.

Eq.~\eqref{qqqq}  can be rewritten in terms of the quadrature fluctuations, which can be cast in the following matrix form	
\begin{align}
	\begin{split}
		\dot{u}(t)=Au(t)+n(t),
	\end{split}
\end{align}
where $u(t)=[\delta X_m(t), \,\, \delta Y_m(t), \,\, \delta X_b(t), \,\, \delta Y_b(t)]^T$ denotes the vector of the quadrature fluctuations, $n(t)=[\sqrt{\kappa_m}X_m^{in}(t),\sqrt{\kappa_m}Y_m^{in}(t),\sqrt{\kappa_b}X_b^{in}(t),\sqrt{\kappa_b}Y_b^{in}(t)]^T$ is the vector of the input noises, and the quadratures are defined as $X_O=\frac{1}{\sqrt{2}}(O+O^\dagger)$ and $Y_O=\frac{i}{\sqrt{2}}(O^\dagger -O)$, and $\delta X_O$ and $\delta Y_O$ are the corresponding fluctuations. Similarly, the associated input noise operators $X_O^{in}$ and $Y_O^{in}$ can be defined. The drift matrix $A$ is given by

\begin{eqnarray}    
	A=\begin{pmatrix}-\frac{\kappa_{m}}{2}&0&0&G_++G_-\\0&-\frac{\kappa_{m}}{2}&G_+-G_-&0\\0&G_+-G_-&-\frac{\kappa_{b}}{2}&0\\G_++G_-&0&0&-\frac{\kappa_{b}}{2}\end{pmatrix}.
\end{eqnarray}
Owing to the linearized dynamics and Gaussian input noises, the state of the system at any time is a two-mode Gaussian state, which is fully characterized by a $4\times4$ covariance matrix (CM) $V$, with its entries defined as $V_{ij}=\langle u_i(t)u_j(t)+u_j(t)u_i(t)\rangle/2$ ($i,j = 1, 2$). {The dynamical CM $V$ can be obtained by solving the equation
	\begin{align}
		\begin{split}
			\dot{V}=AV+VA^T+D,
		\end{split}
\end{align} }
where $D=$ diag$\big[\kappa_m(N_m+\frac{1}{2}),\kappa_m(N_m+\frac{1}{2}),\kappa_b(N_b+\frac{1}{2}),\kappa_b(N_b+\frac{1}{2}) \big]$ is the diffusion matrix and defined via $D_{ij}\delta(t-t')=\langle n_i(t)n_j(t')+n_j(t')n_i(t)\rangle/2$.

\begin{figure}[b]
	\centering
	\includegraphics[width=0.85\linewidth]{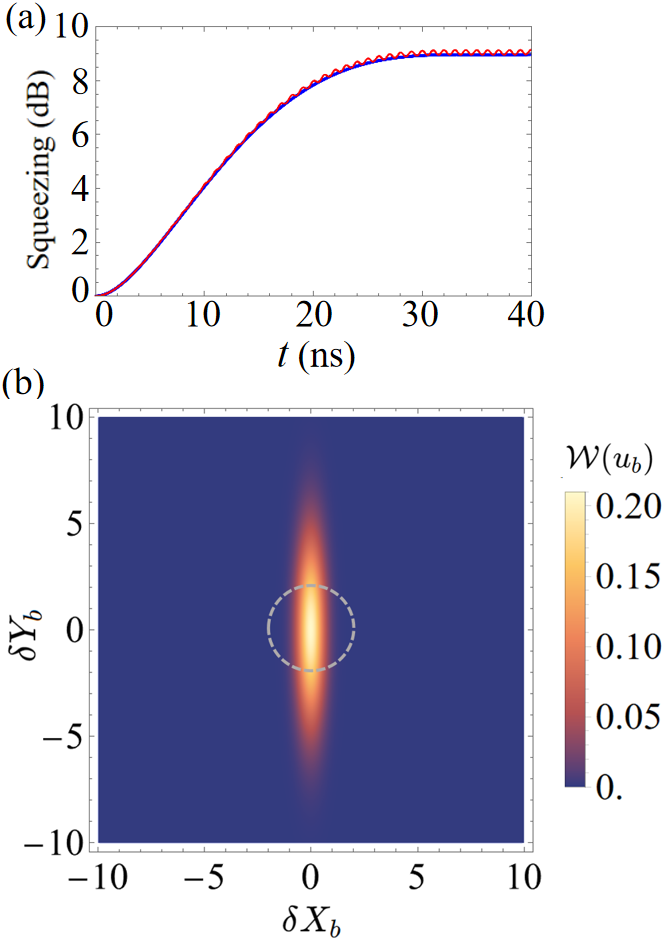}
	\caption{{(a) Degree of squeezing (dB) of the mechanical mode versus time. The blue curve is obtained by using Eq.~\eqref{qqqq} with the RWA, while the thinner red curve is achieved using the full equations Eq.~\eqref{QQQ} without the RWA. (b) Wigner function of the squeezed mechanical mode. The dashed circle corresponds to vacuum fluctuation. The parameters are provided in the main text.}}
	\label{fig2}
\end{figure}

For our two-mode Gaussian state, the CM can be expressed in the form 
\begin{eqnarray}    
	V\equiv \begin{pmatrix}V_m&V_{\rm mb} \\  V_{\rm mb}^T&V_b\end{pmatrix},
\end{eqnarray}
and the Wigner function of the mechanical mode can be achieved via \cite{RS87} 
\begin{align}
	\begin{split}
		W(u_b)=\frac{1}{2\pi \sqrt{{\rm det} V_b}} {\rm exp} \left(-\frac{u^T_b V_b^{-1} u_b}{2} \right),
	\end{split}
\end{align}
with $u_b=(\delta X_b,\delta Y_b)^T$. 

\begin{figure}[t]
	\centering
	\includegraphics[width=0.9\linewidth]{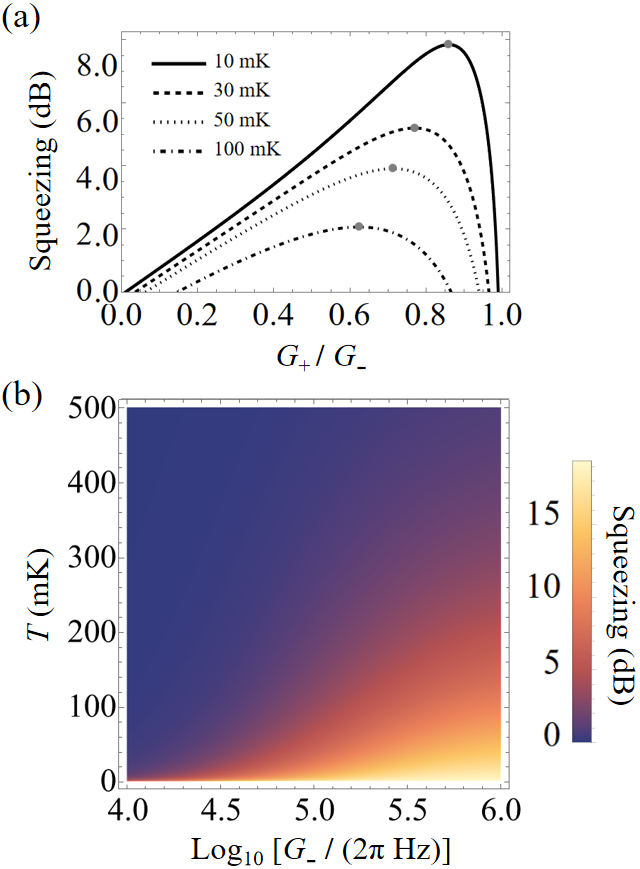}
	\caption{(a) Degree of squeezing (dB) of the mechanical mode versus $G_+/G_-$ for various temperatures. We fix the coupling $G_-/2\pi = 0.1$ MHz and vary $G_+$. (b) Degree of squeezing $S$ (dB) versus bath temperature $T$ and the coupling $G_-$. The coupling $G_+$ is optimized for each value of $G_-$, and the other parameters are as those in Fig.~\ref{fig2}.}
	\label{fig3}
\end{figure}

{Figure~\ref{fig2}(a) shows the degree of squeezing (in units of dB) of the mechanical mode versus time, with and without taking the RWA approximation.  The degree of squeezing is defined as $S =-10 {\rm log}_{10} [V_{\rm min}/V_{\rm vac}]$, with $V_{\rm min}$ being the minimum variance of the mechanical quadrature and $V_{\rm vac}$ being the variance of the vacuum fluctuation. Clearly, the two curves in Fig.~\ref{fig2}(a) almost overlap, indicating that under the parameters listed below, the RWA is a very good approximation.  The mechanical mode evolves into a stationary state with a constant degree of squeezing when the pulses last for $\sim$ 30 ns.  Therefore, this time determines the length of the two microwave pulses used in the first step, i.e., we take $\tau_{\rm sq}=30$~ns.}  We use experimentally feasible parameters \cite{QST23,LPR23,PRAp}: $\omega_m/2\pi=10$~GHz, $\omega_b/2\pi=30$ MHz, {$\kappa_m/2\pi=2$ MHz}, $\kappa_b/2\pi=100$ Hz, and $T=10$ mK. We take $G_-/2\pi=0.1$ MHz and an optimized  $G_+=0.885G_-$, which correspond to the powers of the two drive fields $P_-\simeq 0.36$ mW and $P_+\simeq 0.28$ mW for a $5\times2\times1$ $\mu$m$^3$ YIG micro bridge with $G_0/2\pi=10$ Hz \cite{QST23,LPR23}.    Figure~\ref{fig2}(b) exhibits the corresponding Wigner function of the mechanical squeezed state, which clearly shows that the fluctuation of one mechanical quadrature is significantly reduced below the vacuum fluctuation.

\begin{figure}[t]
	\centering
	\includegraphics[width=0.8\linewidth]{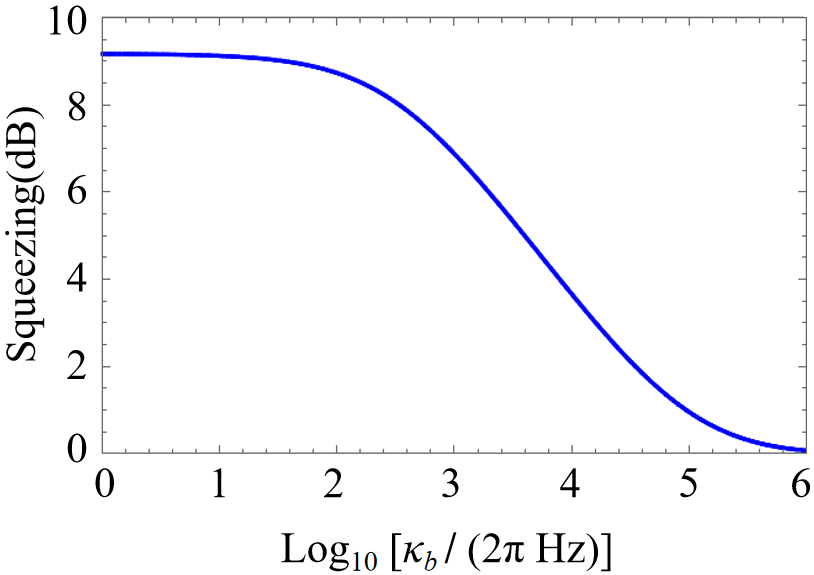}
	\caption{{Degree of squeezing (dB) of the mechanical mode versus mechanical damping rate $\kappa_b$. The other parameters are as those in Fig.~\ref{fig2}}}
	\label{fig4}
\end{figure}

In Fig.~\ref{fig3}(a), we plot the degree of squeezing $S$ as a function of the ratio $G_+/G_-$ for various bath temperatures.
Clearly, there is an optimal ratio $G_+/G_-$, which decreases as the temperature rises. This is due to the fact that, on the one hand,  the difference between the couplings $G_+$ and $G_-$ should be sufficiently large in order to efficiently cool the mechanical mode; on the other hand, almost equal couplings $G_+ \simeq G_-$ is required to have an extremely large squeezing \cite{Jie17}. The trade-off between these two opposite trends results in an optimal ratio $G_+/G_-$. As the temperature rises, a larger cooling rate is needed to cool the mechanical mode with more thermal noise, which thereby leads to a reduced optimal ratio and a reduced degree of squeezing.   In Fig.~\ref{fig3}(b), we plot the degree of squeezing $S$ versus the coupling strength $G_-$ and the temperature $T$. It tells that the maximum degree of squeezing increases with the coupling $G_-$ (with $G_+$ being optimized) and decreases with the temperature, in consistency with the results of Fig.~\ref{fig3}(a) for a fixed $G_-$. 
{Note that the microwave drive pulses might cause heating of the YIG crystal. Nevertheless, the possible heating leads to a temperature rise, whose effect is included in Fig.~\ref{fig3}(b). It shows that even for the temperature being up to hundreds of mK, the degree of squeezing is still considerable. In Fig.~\ref{fig4}, we further analyze the effect of the mechanical damping rate. Clearly, the degree of squeezing remains high, about 7 dB, for $\kappa_b$ being up to 1 kHz, corresponding to a mechanical $Q$ factor of $3\times 10^4$. } 

\section{PHONON SUBTRACTION via optical pulses}
\label{resul}

Once the mechanical mode is prepared in a squeezed state (ideally a squeezed vacuum state, {but practically a squeezed thermal state due to the added thermal noise~\cite{IS20,Tan20,Schwab15}}), we switch off the two microwave pulses. After a short period for magnons to decay, we then send an optical pulse to the optomechanical subsystem to implement the phonon subtraction operation, which we show below can yield a mechanical cat-like state. 

Since the microwave driving fields are turned off, the effective magnomechanical coupling becomes very weak. The system, under an optical driving field, then becomes an effective two-mode system with an optomechanical Hamiltonian, given by
\begin{align}
\begin{split}
	H_{\rm OM}/\hbar=\omega_cc^\dagger c+\omega_bb^\dagger b-g_0c^\dagger c \left(b^\dagger+b \right)-E \left(e^{-i\omega_0t}a^\dagger -\mathrm{H.c.} \right),
\end{split}
\end{align}
where $E=\sqrt{\kappa_cP_0/(\hbar\omega_0)}$ denotes the coupling strength between the optical cavity and the driving field, with $P_0$ ($\omega_0$) being the power (frequency) of the driving field and $\kappa_c$ being the cavity decay rate.

Similarly, we linearize the dynamics following the procedures in Sec.~\ref{sqzing}. When the drive field is red-detuned from the cavity by $\Delta \equiv \omega_c-\omega_0=\omega_b$, and in the resolved sideband limit  $\omega_b \gg \kappa_c$, we derive the following linearized QLEs for the quantum fluctuations $\{\delta c$, $\delta b\}$, which, in the interaction picture with respect to $\hbar\omega_0c^\dagger c$, are given by~\cite{JL21}
\begin{align}\label{qqqqq}
\begin{split}
	\delta \dot{c}&=(-i\Delta-\frac{\kappa_c}{2})\delta c+iG_c\delta b+\sqrt{\kappa_c}c_{in} , \\
	\delta \dot{b}&=(-i\omega_b-\frac{\kappa_b}{2})\delta b+iG_c\delta c+\sqrt{\kappa_b}b_{in},
\end{split}
\end{align}
where $c_{in}$ is the input noise of the cavity, and the effective optomechanical coupling $G_c=g_0 \langle c \rangle$, with $\langle c \rangle=\frac{iE}{i\Delta +\frac{\kappa_c}{2}} \simeq \frac{E}{\omega_b}$. Again, we consider a flattop pulse such that $G_c$ is constant during the pulse. Note that in getting Eq.~\eqref{qqqqq}, we have neglected the frequency shift induced by the optomechanical interaction as it is typically much smaller than $\Delta =\omega_b$.

{We consider the pulse duration to be much shorter than the mechanical life and coherence time, i.e., $\tau_s \ll (N_b \kappa_b)^{-1} \simeq 0.25$ ms, such that the mechanical dissipation and decoherence are negligible during the pulse~\cite{IS20,Tan20,Jie18a}.} The corresponding QLEs, in the interaction picture with respect to $\hbar\omega_c c^\dagger c + \hbar\omega_b b^\dagger b$, are given by
\begin{align}
\begin{split}
	\delta \dot{c}&=-\frac{\kappa_c}{2}\delta c+iG_c\delta b+\sqrt{\kappa_c}c_{in}, \\
	\delta \dot{b}&=iG_c \delta c,
\end{split}
\end{align}
The pulse is sufficiently weak, yielding $G_c\ll \kappa_c$, which allows us to adiabatically eliminate the cavity, and obtain  $\delta c=i\frac{2G_c}{\kappa_c}\delta b+\frac{2}{\sqrt{\kappa_c}}c_{in}$. By using the input-output relation  $c_{out}=\sqrt{\kappa_c} \delta c -c_{in}$, we get 
\begin{align}
\begin{split}
	c_{out} &=i\sqrt{2G} \delta b + c_{in}, \\
	\delta \dot{b}&=-G \delta b+i \sqrt{2G} c_{in},
\end{split}
\end{align}
where $G=2G_c^2/\kappa_c$.  By further introducing the temporal modes associated with the pulse \cite{Hammerer}
\begin{align}
\begin{split}
	C_{in}(t)&=\sqrt{\frac{2G}{e^{2Gt}-1}}\int_0^tdt'e^{Gt'}c_{in}(t'), \\
	C_{out}(t)&=\sqrt{\frac{2G}{1-e^{-2Gt}}}\int_0^tdt'e^{-Gt'}c_{out}(t'),
\end{split}
\end{align}
which satisfy the commutation relation $[C_k,C_k^\dagger]=1$ ($k=in, out$), we achieve
\begin{align}\label{ccbb}
\begin{split}
	C_{out}(t)&=e^{-Gt}C_{in}(t)+i\sqrt{1-e^{-2Gt}}\delta b(0) , \\
	\delta b(t)&=e^{-Gt} \delta b(0)+i\sqrt{1-e^{-2Gt}}C_{in}(t).
\end{split}
\end{align}
From Eq.~\eqref{ccbb}, we can extract a propagator $U(t)$, satisfying $C_{out}(t)= U(t)^\dagger C_{in}(t) U(t)$ and  $\delta b(t)=  U(t)^\dagger  \delta b(0) U(t)$, given by \cite{JL21}
\begin{align}\label{UUU}
\begin{split}
	U=e^{i\tan\theta C_{in}^\dagger b}\cos\theta ^{-(C_{in}^\dagger C_{in}-b^\dagger b)}e^{-i\tan\theta C_{in} b^\dagger}
\end{split}
\end{align}
where $ \cos \theta\equiv e^{-Gt}$ and $\tan\theta\equiv\sqrt{e^{2Gt}-1}$.   

We first analyze the ideal case where the mechanical mode is prepared in a squeezed vacuum state $|\xi\rangle_b=S(\xi)|0\rangle_b$, with $S(\xi)= e^{\frac{1}{2} (\xi b^{\dagger 2} -\xi^* b^{2})}$ being the squeezing operator and $\xi = re^{i\phi}$, with $r$ being the squeezing parameter and $\phi$ the phase angle, i.e.,
\begin{align}\label{1888}
\begin{split}
	|\xi\rangle_b&=\frac{1}{\sqrt{\cosh r}}\sum_{n=0}^{\infty}(e^{i\phi}\tanh r)^n\frac{\sqrt{(2n)!}}{2^nn!}|2n\rangle_b \\
	&=\sqrt[4]{1-\tanh^2r}\sum_{n=0}^{\infty}C_{n}(\tanh r)^{n}|2n\rangle_b,
\end{split}
\end{align}
with $C_{n} = \left(\frac{e^{i\phi}}{2} \right)^n\frac{\sqrt{(2n)!}}{n!}$. {As will be shown later, the analysis of this ideal  squeezed vacuum pure state is important, because it allows us to see more clearly how the squeezing parameter $r$ evolves due to the pulse interaction}. 
The function of the pulse is equivalent to applying the propagator $U(t)$ onto the initial state $|0\rangle_c|\xi\rangle_b$, where $|0\rangle_c$ denotes the cavity is in vacuum. At the end of the pulse, the system evolves to be	
\begin{align}\label{eeee}
\begin{split}
	|\psi \rangle_{c,b} &= U|0\rangle_c|\xi\rangle_b=\sum_{n=0}^{\infty}\frac{(i\tan \theta)^\textit{n}}{n!} \left(C_{in}^\dagger b \right)^n|0\rangle_c|\xi'\rangle_b\\
	&\approx|0\rangle_c|\xi'\rangle_b+i\tan \theta|1\rangle_c \left(\textit{b}|\xi'\rangle_\textit{b} \right)-\frac{\tan^2 \theta}{\sqrt{2}}|2\rangle_\textit{c}\left(\textit{b}^2|\xi'\rangle_\textit{b} \right)\\
	&{-i\frac{\tan^3 \theta}{\sqrt{6}}|3\rangle_c\left(b^3|\xi'\rangle_b\right)},
\end{split}
\end{align}
where
\begin{align}\label{2000}
\begin{split}
	|\xi'\rangle_b=\sqrt[4]{1-\left(\tanh r \, \cos^2 \theta \right)^2}\sum_{n=0}^{\infty}C_{n}\left( \tanh r \, \cos^2 \theta \right)^{n}|2n\rangle_b.
\end{split}
\end{align}
In getting the last line of Eq.~\eqref{eeee}, we have omitted higher-order terms with $n >2$, which is a good approximation when $\tan^2 \theta \ll 1$, i.e., $ Gt \ll 1$. Equation~\eqref{eeee} indicates that the subtraction of a single phonon  $b|\xi'\rangle_{b}$ can be realized when a single anti-Stokes photon in the cavity output field is detected, and the successful probability is approximately $\tan^2 \theta$. Similarly, the subtraction of two {(three)} phonons  $b^2|\xi'\rangle_{b}$ {($b^3|\xi'\rangle_{b}$)} can also be achieved, with a much smaller probability $\sim \tan^4 \theta$ {($\tan^6 \theta$)}, conditioned on the detection of a two {(three)}-photon state.  We note that due to the pulse interaction, the squeezing $r$ of the squeezed vacuum state is slightly reduced,  because $\tanh  r \rightarrow \tanh  r \,\cos^2 \theta$ as seen in Eq.~\eqref{2000}.

{In a practical situation, we prepare a squeezed thermal mixed state, which is reflected in the CM $V_b$ of the mechanical mode obtained in Sec.~\ref{sqzing}, from which two key characteristic parameters of the squeezed thermal state are extracted, i.e., the squeezing parameter $r=1.23$, and the residual phonon number $\bar{n}=0.014$ (see Appendix for details).} To be compatible with the propagator $U(t)$, we transform the CM $V_b$ into a density matrix $\rho_{in,b}$ (Appendix), which acts as the initial state before the optical pulse is applied. Combining the vacuum state of the cavity, the initial state of the optomechanical system is then $\rho_{in}=\rho_{in,b}\otimes|0\rangle\langle0|_c$, which, at the end of the pulse, evolves into the state $\rho_{b,c}=U\rho_{in}U^\dagger$, given by \cite{MD97,MB94}
\begin{align}\label{xxxx}
\begin{split}
	\rho_{b,c}&=\sum_{n=0}^{\infty}\sum_{m=0}^{\infty} \frac{e^{-i(m-n)\pi/2}}{\sqrt{n!m!}}(-1)^{m+n} |\tan\theta|^{m+n}\\
	&\times b^m|\cos\theta|^{b^\dagger b}\rho_{in,b} |\cos\theta|^{b^\dagger b}\textit{b}^{\dagger n}\otimes|m\rangle\langle n|_\textit{c}.
\end{split}
\end{align}

We consider the situation where a $n$-photon state is detected in the output field, i.e., taking $m=n$ in Eq.~\eqref{xxxx}, which gives rise to the following state
\begin{align}\label{zzzz}
\begin{split}
\rho_{b,c}&=\sum^\infty_{n = 0}\frac{(\tan\theta)^{2n}}{n!}b^n |\cos\theta|^{b^{\dagger}b} \rho_{in,b} |\cos\theta|^{b^\dagger b}b^{\dagger n}\otimes|\textit{n}\rangle\langle \textit{n}|_\textit{c} \\
&\approx|\cos\theta|^{b^\dagger b}  \rho_{in,b} |\cos\theta|^{b^\dagger b}\otimes|0\rangle\langle 0|_\textit{c}\\
&+\tan^2 \theta \, b|\cos\theta|^{b^\dagger b}\rho_{in,b}|\cos\theta|^{b^\dagger b}b^{\dagger}\otimes|1\rangle\langle 1|_\textit{c}\\
&+\frac{\tan^4 \theta}{2}b^2|\cos\theta|^{b^\dagger b}\rho_{in,b}|\cos\theta|^{b^\dagger b}b^{\dagger 2}\otimes|2\rangle\langle 2|_\textit{c}\\
&+{\frac{\tan^6 \theta}{6}b^3|\cos \theta|^{b^\dagger b}\rho_{in,b}|\cos \theta|^{b^\dagger b}b^{\dagger 3}\otimes|3\rangle\langle3|_\textit{c}},
\end{split}
\end{align}
{where higher-order terms with {$n >3$} are neglected because of their extremely small probabilities. The above equation indicates that $k$-phonon can be subtracted from the initial squeezed thermal state, i.e., $b^k \rho_{in,b} b^{\dagger k}$ ($k=1,2,3$), conditioned on the detection of a $k$-photon state in the cavity output.} Note that the slightly reduced $r$ revealed in Eq.~\eqref{2000} is not visible because a general mixed state $\rho_{in,b}$ is adopted in Eq.~\eqref{zzzz}.

\begin{figure*}[t]
\centering
\includegraphics[width=0.93\linewidth]{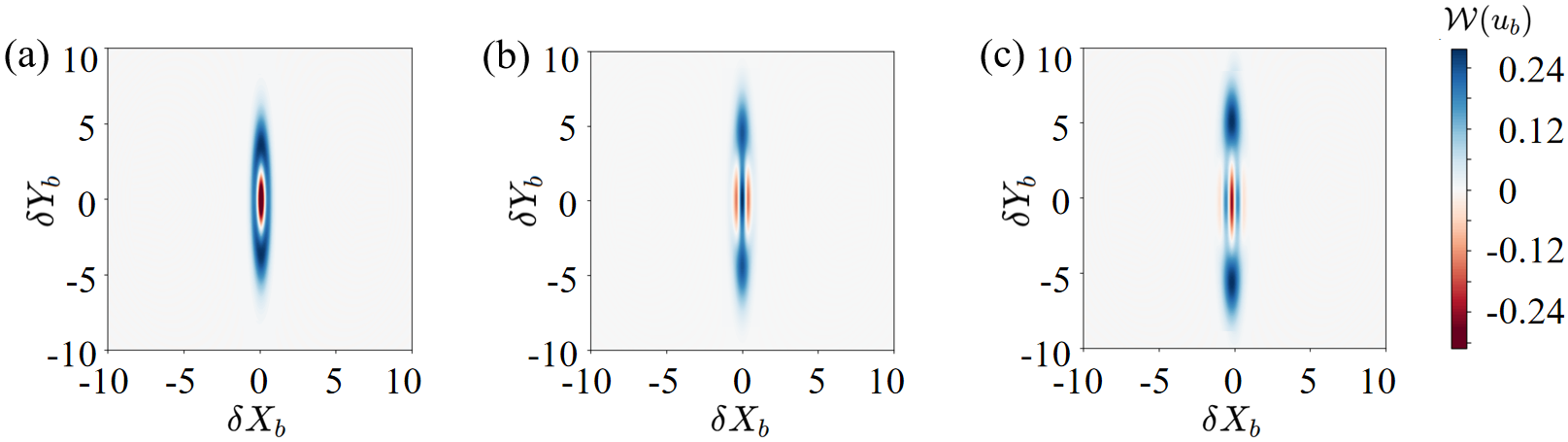}
\caption{Wigner function of (a) the single-phonon (b) the two-phonon {(c) the three-phonon} subtracted squeezed mechanical state. The parameters are provided in the text.}
\label{fig5}
\end{figure*}

In Fig.~\ref{fig5}, we plot the Wigner function of the single/two{/three}-phonon-subtracted squeezed mechanical states. The phonon subtraction leads to the negativity of the Wigner function, demonstrating the nonclassical nature of the state. Moreover, the Wigner function exhibits a coherent superposition feature with the interference fringes shown around the origin of the phase space, which resembles that of the Schr\"odinger's cat state. {For this reason, it is called a mechanical cat-like state.}  We take the following parameters: optical wavelength $\lambda_0=1550$ nm, {$\kappa_c/2\pi=8$ MHz, $g_0/2\pi=100$ Hz, pulse duration $\tau_s=10$ ns, and power $P_0 = 60$ nW}, which yield $G/2\pi = 1.8$ kHz and $\tan \theta \approx 0.11$.  {The probability of subtracting a single phonon, two phonons, and three phonons are on the order of $\sim 10^{-2}$, $10^{-4}$, and $10^{-6}$, respectively.}

{To see how close our prepared state is to the ideal cat state, we calculate the fidelity with the cat state, which is defined as  
\begin{align}
\begin{split}
	{F_{\pm}= \sqrt{ \leftindex_{\pm}\langle \psi_{\alpha_m}|\rho_b|\psi_{\alpha_m}\rangle_{\pm}} },
\end{split}
\end{align}
where $\rho_b= \leftindex_c\langle k|\rho_{b,c}|k\rangle _c/N_0$ (with a normalization factor $N_0^{-1}$) 
is the $k$-phonon-subtracted squeezed thermal state, and $|\psi_{\alpha_m}\rangle_{\pm}$ is the even or odd cat state with the amplitude $\alpha_m$, i.e. $|\psi_{\alpha_m}\rangle_{\pm}=(|\alpha_m\rangle \pm |-\alpha_m\rangle)/\sqrt{1\pm e^{-2\alpha_m^2}}$. 
The fidelity of the three cat-like states in Fig.~\ref{fig5} are 87$\%$, 91$\%$, and 94$\%$, respectively. The fidelity increases when more phonons are subtracted from the squeezed state, which corresponds to a smaller successful probability and thus longer measurement time.  Note that for the single- and three-phonon subtracted states we compare with the odd cat state, whereas for the two-phonon subtracted state, we compare with the even cat state. For the calculation of fidelity of each state, an optimal $\alpha_m$ is used, which is 3.15, 5.03, and 5.77, respectively, for $k=1,2,3$.
In addition, we calculate the macroscopicity of the cat-like states, a measure to quantify macroscopic quantum superpositions, which is defined as~\cite{CW11}
\begin{align}
\begin{split}
	{\mathcal{I}=-\frac{\pi}{2}\iint W(x,p)\left(\frac{\partial^2}{\partial x}+\frac{\partial^2}{\partial p}+2\right)W(x,p) {\rm d}x {\rm d}p}.
\end{split}
\end{align}
The macroscopicity of the three cat-like states in Fig.~\ref{fig5} are 3.4, 4.7, and 5.7, respectively. }

\begin{figure}
\centering
\includegraphics[width=1\linewidth]{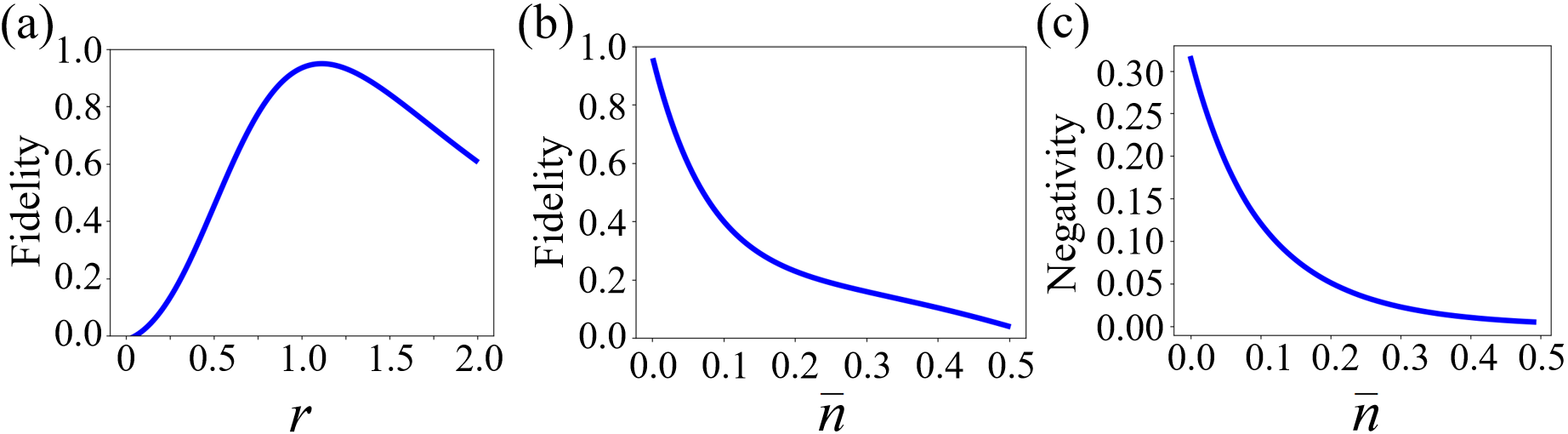}
\caption{{The fidelity of the two-phonon subtracted squeezed thermal state to the cat state versus (a) the squeezing parameter $r$ and (b) the residual phonon number $\bar{n}$. (c) The corresponding negativity of the Wigner function versus $\bar{n}$. We fix $\bar{n} = 0.014$ in (a) and $r = 1.23$ in (b) and (c).  The other parameters are the same as in Fig.~\ref{fig5}. }}
\label{fig6}
\end{figure}

{To further analyze the effects of some key parameters on our cat-like states, we plot in Figs.~\ref{fig6}(a) and \ref{fig6}(b) the fidelity of the two-phonon subtracted squeezed state (to the cat state with $\alpha_m=$ 5.03) versus the two characteristic parameters of the squeezed thermal state, i.e., the squeezing parameter $r$ and the residual phonon number $\bar{n}$, respectively. Clearly, the fidelity increases with the squeezing when $r<1$, but a too large $r$ is not helpful as it reduces the fidelity with the cat state due to the distortion of the two superposed wave packets.   The fidelity decreases with the residual phonon number $\bar{n}$, indicating that the optimal situation corresponds to the squeezed vacuum state ($\bar{n}=0$) since the ideal cat state $|\psi_{\alpha_m}\rangle_{\pm}$ is a pure state.  In Fig.~\ref{fig6}(c), we further plot the negativity of the Wigner function of the two-phonon subtracted squeezed state versus $\bar{n}$, which is defined as the volume of the negative Wigner distributions ${\mathcal{N}=\frac{1}{2}(\iint |W(x,p)|dx dp -1)}$~\cite{AK04}.  Since the negativity only appears in the interference region and the Wigner becomes all non-negative when there is no interference. In this sense, the negativity can be used to infer the interference visibility of the cat-like state.  Figure~\ref{fig6}(c) shows that the interference visibility, as well as the nonclassicality, reduces with $\bar{n}$, and the optimal situation also corresponds to a vanishing $\bar{n}\to0$.  }

{Finally, when the mechanical motion is prepared in the cat-like state, another red-detuned optical pulse with duration $\tau_r$ is sent to the cavity to map the mechanical state to the cavity output field~\cite{Tan20,Jie18a}, from which the tomography of the mechanical state can be subsequently implemented. }

{Compared with the existing protocols using a pure optomechanical system~\cite{IS20,Tan20}, the advantage of our protocol employing a magnomechanical system in the first step (squeezing mechanical motion) lies in the fact that the magnon dissipation rate of the YIG crystal is the lowest among all magnetic materials~\cite{Chumak}: $\kappa_m$ is around 1 MHz, which is much smaller than the mechanical frequency of both a YIG sphere~\cite{Zuo} and a YIG micro bridge~\cite{PRAp}. Thus, for the magnomechanical system, the resolved-sideband limit $\kappa_m \ll \omega_b$ is always well satisfied, which remains challenging for many optomechanical configurations~\cite{MA14}. The resolved-sideband limit is required for the reservoir-engineering method to be effective and competent to generate strong mechanical squeezing. Moreover, a smaller magnon dissipation rate leads to a larger cooperativity, which is a key parameter to achieve both strong squeezing and high purity of the mechanical state~\cite{IS20}. }

{We now analyze the effects of other experimental imperfections on our protocol.  For photon detection and transmission loss, the photon loss reduces the heralding probability~\cite{IS20,JL20}. For the anti-Stoke photon that fails to be detected (due to finite detection efficiency), it is not recorded and thus this event is discarded, so lower detection efficiency implies longer measurement time. For dark counts and false heralding, it means that the detector clicks while the anti-Stoke scattering is not actually activated. This event is recorded, but the phonon is not subtracted, and thus the mechanical mode remains in the squeezed state.  This can be discriminated by subsequent tomography, where the true heralding corresponds to negative Wigner around the origin of phase space (Fig.~\ref{fig5}), while the false heralding gives a positive Wigner at the origin (Fig.~\ref{fig2}(b)).   }

\section{Conclusions}
\label{conc}

We present a two-step pulse scheme to prepare a cat-like state of mechanical motion using an OMM system. By applying a two-tone microwave drive field to the magnomechanical system, the mechanical mode is prepared in a squeezed thermal state in the first step. By further sending a weak red-detuned optical pulse to the optomechanical cavity in the second step, $k$ phonons can be subtracted from the squeezed state conditioned on the detection of $k$ anti-Stokes photons from the cavity output field, which results in a mechanical cat-like state. More phonons subtracted from the squeezed state correspond to a higher fidelity to the cat state, but a lower successful probability. The work may find applications in macroscopic quantum studies, quantum sensing, and the test of collapse models.

\section*{ACKNOWLEDGMENTS}

This work was supported by Zhejiang Provincial Natural Science Foundation of China (Grant No. LR25A050001), National Natural Science Foundation of China (Grants No. 12474365 and No. 92265202), and National Key Research and Development Program of China (Grants No. 2024YFA1408900 and No. 2022YFA1405200).

\setcounter{figure}{0}
\setcounter{equation}{0}
\setcounter{table}{0}
\renewcommand\theequation{A\arabic{equation}}
\renewcommand\thefigure{A\arabic{figure}}
\renewcommand\thetable{A\arabic{table}}

\section*{APPENDIX: TRANSFORMATION FROM COVARIANCE MATRIX TO DENSITY MATRIX}\label{appA}

In Sec.~\ref{sqzing}, we prepare a squeezed thermal mechanical state characterized by a $2\times2$ CM $V_b$, which acts as the initial state $\rho_{in,b}$ for the subsequent operation of phonon subtraction. To exploit the derived propagator $U(t)$ associated with the pulse, we need to transform the form of the mechanical state from the CM to the density matrix. The squeezed thermal state can be expressed as $\rho_{in,b}=S(\xi)\rho_{th} (\bar{n})S^\dagger(\xi)$, where $S(\xi)$ is the squeezing operator, $\xi = re^{i\phi}$, and $\rho_{th}=\Sigma_nP_n|n\rangle\langle n|$ is the thermal state, $P_n=\bar{n}^n/(1+\bar{n})^{n+1}$, with $\bar{n}$ being the mean thermal phonon number. The characteristic parameters $r$, $\phi$, and $\bar{n}$ can be extracted from the CM $V_b$ via~\cite{VG06}
\begin{align}
\begin{split}
r&=\frac{1}{2}{\rm arcosh} \, \frac{{\rm Tr}\, V_b}{2\sqrt{{\rm det}\, V_b}},\\
\bar{n} &=\frac{\sqrt{{\rm det} \, V_b}-1}{2},\\
\tan\phi&=\frac{2V_{12}}{V_{11}-V_{22}}, 
\end{split}
\end{align}
with $V_{j,k}$ ($j,k =1,2$) being the elements of $V_b$. For the mechanical state with {$V_b=\{ \{0.046 ,0.141\} , \{0.141 ,6.283\}\}$ prepared in Sec.~\ref{sqzing}, we get $r=1.23$, $\phi=-0.05$, and $\bar{n}=0.014$. }


\begin{thebibliography}{99}

\bibitem{CM96}
C. Monroe, D. M. Meekhof, B. E. King, and D. J. Wineland, Science {\bf272}, 1131 (1996).

\bibitem{DL05}
D. Leibfried, E. Knill, S. Seidelin, J. Britton, R. B. Blakestad, J. Chiaverini, D. B. Hume, W. M. Itano, J. D. Jost, C. Langer, R. Ozeri, R. Reichle, and D. J. Wineland, Nature {\bf438}, 639 (2005).

\bibitem{BV13}
B. Vlastakis, G. Kirchmair, Z. Leghtas, S. E. Nigg, L. Frunzio, S. M. Girvin, M. Mirrahimi, M. H. Devoret, and R. J. Schoelkopf, Science {\bf342}, 607 (2013).

\bibitem{KH15}
K. Huang, H. L. Jeannic, J. Ruaudel, V. B. Verma, M. D. Shaw, F. Marsili, S. W. Nam, E. Wu, H. Zeng, Y.-C. Jeong, R. Filip, O. Morin, and J. Laurat, Phys. Rev. Lett. {\bf115}, 023602 (2015).

\bibitem{AO06}
A. Ourjoumtsev, R. Tualle-Brouri, J. Laurat, and P. Grangier, Science {\bf312}, 83 (2006).

\bibitem{JSN06}
J. S. Neergaard-Nielsen, B. M. Nielsen, C. Hettich, K. Molmer, and E. S. Polzik, Phys. Rev. Lett. {\bf97}, 083604 (2006).

\bibitem{AO19}
A. Omran {\it et al}., Science {\bf365}, 570 (2019).

\bibitem{BH19}
B. Hacker, S. Welte, S. Daiss, A. Shaukat, S. Ritter, L. Li, and G. Rempe, Nat. Photon, {\bf13}, 110 (2019).

\bibitem{MB23}
M. Bild, M. Fadel, Y. Yang, U. V. L\"upke, P. Martin, A. Bruno, and Y. Chu, Science {\bf380}, 274 (2023).

\bibitem{IS20}
I. Shomroni, L. Qiu, and T. J. Kippenberg, Phys. Rev. A {\bf101}, 033812 (2020).

\bibitem{Tan20}
H. Zhan, G. Li, and H. Tan,  Phys. Rev. A {\bf101}, 063834 (2020).

\bibitem{MA14}
M. Aspelmeyer, T. J. Kippenberg, and F. Marquardt, Rev. Mod. Phys. {\bf86}, 1391 (2014).

\bibitem{Zuo}
X. Zuo, Z. Fan, H. Qian, M. Ding, H. Tan, H. Xiong, and J. Li, New J. Phys. {\bf26}, 031201 (2024).

\bibitem{JL18}
J. Li, S.-Y. Zhu, and G. S. Agarwal, Phys. Rev. Lett. {\bf121} 121, 203601 (2018).


\bibitem{JL19b}
J. Li and S.-Y. Zhu, New J. Phys. {\bf21}, 085001 (2019).

\bibitem{Jing24}
Q. Zhang, J. Wang, T.-X. Lu, R. Huang, F. Nori, and H. Jing, Sci. China-Phys. Mech. Astron. {\bf 67}, 100313 (2024).

\bibitem{RCS22}
R.-C. Shen, J. Li, Z.-Y. Fan, Y.-P. Wang, and J.-Q. You, Phys. Rev. Lett. {\bf129}, 123601 (2022).

\bibitem{Xiong23}
H. Xiong, Fundam. Res.  {\bf 3}, 8 (2023).

\bibitem{Dong23}
G.-T. Xu, M. Zhang, Y. Wang, Z. Shen, G.-C. Guo, and C.-H. Dong, Phys. Rev. Lett. {\bf 131}, 243601 (2023).


\bibitem{Fan}
Z. Fan, R. Shen, Y. Wang, J. Li, and J. Q. You. Phys. Rev. A {\bf 105}, 033507 (2022).

\bibitem{QST23}
Z. Fan, H. Qian, and J. Li, Quantum Sci. Technol. {\bf 8}, 015014 (2023).

\bibitem{Dong22}
Z. Shen, G.-T. Xu, M. Zhang, Y.-L. Zhang, Y. Wang, C.-Z. Chai, C.-L. Zou, G.-C. Guo, and C.-H. Dong, Phys. Rev. Lett. {\bf129}, 243601 (2022).


\bibitem{LPR23}
Z. Fan, L. Qiu, S. Gr\"oblacher, and J. Li, Laser Photonics Rev. {\bf17}, 2200866 (2023).

\bibitem{LPR25}
H. Li, Z. Fan, H. Zhu, S. Gr\"oblacher, and J. Li, Laser Photonics Rev. {\bf19}, 2401348 (2025).

\bibitem{Xiong25}
M.-Y. Liu, Y. Gong, J. Chen, Y.-W. Wang, and W. Xiong, Chinese Phys. B {\bf 34}, 057202 (2025).

\bibitem{Zhang25}
C. Zhang, X. Liu, L. Gao, R. Yang, J. Zhang, and T. Zhang, Opt. Express {\bf 33}, 33330 (2025).

\bibitem{FanPRA23}
Z. Fan, H. Qian, X. Zuo, and J. Li, Phys. Rev. A {\bf 108}, 023501 (2023).

\bibitem{Di24}
K. Di, X. Wang, H. Xia, Y. Zhao, Y. Liu, A. Cheng, and J. Du, Opt. Lett. {\bf 49}, 2878 (2024).



\bibitem{MD97}
M. Dakna, T. Anhut, T. Opatrny, L. Kn\"oll, and D.-G. Welsch, Phys. Rev. A, {\bf55}, 3184 (1997).

\bibitem{AB07}
A. Biswas and G. S. Agarwal, Phys. Rev. A {\bf75}, 032104 (2007).

\bibitem{TJ16}
T. J. Milburn1, M. S. Kim, and M. R. Vanner, Phys. Rev. A {\bf93}, 053818 (2016).

\bibitem{Kittel}
C. Kittel, Phys. Rev. {\bf 73}, 155 (1948).







\bibitem{PRAp}
F. Heyroth, C. Hauser, P. Trempler, P. Geyer, F. Syrowatka, R. Dreyer, S. G. Ebbinghaus, G. Woltersdorf, and G. Schmidt, Phys. Rev. Appl. {\bf 12}, 054031 (2019).


\bibitem{SG}
S. Gr\"oblacher, K. Hammerer, M. R. Vanner, and M. Aspelmeyer, Nature {\bf 460}, 724 (2009).

\bibitem{Schwab15}
E. E. Wollman {\it et al.}, Science {\bf 349}, 952 (2015).

\bibitem{Jie18a}
J. Li, S. Gr\"oblacher, S.-Y. Zhu, and G. S. Agarwal, Phys. Rev. A {\bf 98}, 011801(R) (2018).

\bibitem{YDW13}
Y.-D. Wang and A. A. Clerk, Phys. Rev. Lett. {\bf 110}, 253601 (2013).

\bibitem{HT13}
H. Tan, G. Li, and P. Meystre, Phys. Rev. A {\bf 87}, 033829 (2013).

\bibitem{JL15}
J. Li, I. M. Haghighi, N. Malossi, S. Zippilli, and D. Vitali, New J. Phys. {\bf 17}, 103037 (2015).




\bibitem{RS87}
R. Simon, E C G. Sudarshan, and N. Mukunda, Phys. Rev. A {\bf 36}, 3868 (1987).


\bibitem{Jie17}
J. Li, G. Li, S. Zippilli, D. Vitali, and T. Zhang, Phys. Rev. A {\bf 95}, 043819 (2017).

\bibitem{JL21}
J. Li, Y. P. Wang, W. J. Wu, S. Y. Zhu, and J. Q. You, PRX Quantum, {\bf2}, 040344 (2021).

\bibitem{Hammerer}
S. G. Hofer, W. Wieczorek, M. Aspelmeyer, and K. Hammerer, Phys. Rev. A {\bf 84}, 052327 (2011).



\bibitem{MB94}
M. Ban, Phys. Rev. A, {\bf49}, 5078 (1994).

\bibitem{CW11}
C.-W. Lee and H. Jeong, Phys. Rev. Lett. {\bf 106}, 220401 (2011).

\bibitem{AK04}
A. Kenfack and K. Zyczkowski, J. Opt. B {\bf 6}, 396 (2004).

\bibitem{Chumak}
R. O. Serha, C. Dubs, A. V. Chumak, Magnetic Materials for Quantum Magnonics, arXiv:2510.09331v1.


\bibitem{JL20}
J. Li, A.  Wallucks, R. Benevides, N. Fiaschi, B. Hensen, T. P. Alegre and S. Gröblacher, Phys. Rev. A {\bf 102}, 032402 (2020).


\bibitem{VG06}
V. Giovannetti, S. Lloyd, and L. Maccone, Phys. Rev. Lett. {\bf 96}, 010401 (2006).









\end{thebibliography}
\end{document}